# DSL development based on target meta-models. Using AST transformations for automating semantic analysis in a textual DSL framework

Andrey Breslav

St. Petersburg State University of Information Technology, Mechanics and Optics
abreslav@gmail.com

**Abstract.** This paper describes an approach to creating textual syntax for Domain-Specific Languages (DSL). We consider target meta-model to be the main artifact and hence to be developed first. The key idea is to represent analysis of textual syntax as a sequence of transformations. This is made by explicit operations on abstract syntax trees (ATS), for which a simple language is proposed. Text-to-model transformation is divided into two parts: text-to-AST (developed by openArchitectureWare [1]) and AST-to-model (proposed by this paper). Our approach simplifies semantic analysis and helps to generate as much as possible.

**Keywords:** AST DSL textual syntax transformation text-to-model model-to-text

## 1 Introduction

Since a domain-specific language (DSL) is used to express some particular notions within the problem domain, we consider those notions being the main part of the language. Their structure can be described with some meta-model. Let us call this a *target meta-model* because everything around it aims to build some model in this meta-model (a *target model*).[1]

As opposed to it, a DSL has some *syntax* (*concrete syntax* to be more precise) which helps users input the data. It can be textual (as for most programming languages), XML/XMI, diagram (as for UML and many other modeling notations), tree-like (as default EMF editors [2]), meta-programming (like MPS [3] or Kermeta [4]) or anything else imaginable. Concrete syntax representation is translated (*transformed*, in model terms) to a target model. Such transformations are named *syntax-to-model* (*text-to-model* for textual syntax).

---

[1] In classical compiler theory, the closest thing is that *internal representation* of the program being a result of syntax and semantics analysis.

In this paper, we discuss textual syntax. It is described by some (usually context-free) grammar. NOTE: here the *language* (a DSL) is not considered to be just a set of strings defined by the grammar, but mainly a set of possible target models.

### 1.1 Example: CSS

For example, let us have a look at CSS. A CSS style sheet consists of selectors and property-value pairs for these selectors. Selectors could be tag names, possibly specified by classes, ids, nesting specifications etc. This is target meta-model description for CSS. Browsers use CSS target models to apply styles to web pages.

On the other hand, there is a well-known concrete textual syntax involving dot-name notation for classes, sharp-name for ids, braces for property area, etc. This syntax is used by people who write CSS documents.

Concrete syntax for CSS allows grouping properties under one selector or specifying it separately:

```
.some { border-with: 2px; border-color: red }
```

or

```
.some { border-with: 2px }
.some { border-color: red }
```

Target meta-model does not care about it: all the properties are attached to one selector regardless to whether they were grouped or not.

Another strong difference is reference representation. Concrete syntax makes references using textual names (*"some"* to reference corresponding class) and target meta-model uses object references. Thus, some *lookup actions* are needed to transform concrete syntax to target model. These actions look for objects that correspond to name references from textual representations; they are the most valuable part of semantic analysis (excluding validation).

### 1.2 Translation process

A language processor may perform a text-to-model transformation (translation) directly: parser invokes semantic action during the parsing process. This, possibly, shortens translation time but makes the process less modular and far less formal. Lack of formality affects difficulties in automated compiler construction.

Using an abstract syntax tree (AST) as an intermediate artifact may help with this issue. A text-to-model transformation is divided into text-to-AST and AST-to-model parts. The former hold all the syntax-related actions and the latter hold everything about semantic analysis. Such a modular architecture improves capabilities of DSL framework generators.

## 1.3 Related work

Since DSL frameworks recently became rather popular research and development area, there are several ones aimed to develop textual syntax.

The most noticeable and well supported is openArchitectureWare's *xText* [5] framework that allows one to create a DSL infrastructure (including parser and Eclipse-based editor with syntax highlighting, code completion and error markers) by providing rather simple grammar-based definition.

Another framework to be mentioned here is *Guide/Gymnast* [6] developed in IBM and very little known because of almost no documentation or description (we discovered it by accident, because it is included into AlphaWorks' Emfatic bundle).

These frameworks both use EMF and ANTLR [7]. They both just build AST.

**EMF.** Eclipse Modeling Framework (EMF) [2] as meta-model framework. Here we note some key features of it since we use it widely below.

A self-describing meta-model in EMF is called ECore. Main classes there are EDataType (for primitive types), EClass (for classes), EStructuralFeature (for attributes and associations), EReference (for associations), EAttribute (for attributes of primitive types).

Any EMF model has tree-like structure. There is a single root element and all the other elements are contained by the root explicitly or through other elements. Any element except the root is contained by some other element. So EReferences can be containment or cross (which means non-containment), in MOF-like terms, associations are aggregative or not.

**Emfatic.** Default syntax for EMF is tree viewer based editors. Since it is not suitable for the paper, we use a third-party language called Emfatic [8] to describe EMF models. It's syntax is mostly intuitive. The only thing that needs to be mentioned is that containment references are marked with **val** keyword and cross-references are marked with **ref** keyword. Emfatic is used in section 3.

**xText.** As it plays rather important role in our approach, we describe xText shortly here.

The main idea is to create a grammar language that allows building not only the parser but also a text-to-AST transformation. The AST meta-model is described in EMF terms.

Main principle is that a grammar rule having non-terminal X on the left side defines an AST class X. The right side of such a rule refers to other rules by assigning them to X's features. This implicitly defines features' types.

Nothing but such a grammar has to be provided to define AST meta-model and text-to-AST transformation. Section 4.3 gives some examples of xText usage.

This tool is syntax-centric. Its main goal is to build parser that produces AST, not target model. So it does nothing about AST-to-target transformations and lookup (xText allows to perform some semantic analysis but only through constraints checking, it also has connection with *xTend* [9] transformation language, but all the transformations must be written manually) and the concrete syntax grammar is the main artifact they operate on.

### 1.6 Our goal

Develop principles and a framework allowing automatically generating a textual DSL infrastructure concentrating on the target meta-model, not the concrete syntax. We understand that no concrete syntax might be developed without a grammar, but the grammar should not be the process's entry point and the main determining artifact, but the target meta-model should.

## 2 Approach

Since we consider target meta-model to be the main artifact, a DSL development should start with target meta-model definition. This is done manually.

When the meta-model is defined, we start moving towards concrete syntax and the first thing to do here is define AST meta-model (classes of AST nodes). As described below, we propose rather simple transformation language to transform target meta-model to AST meta-model. The main issue here is to translate references to have textual form in AST.

The next step is to define the grammar for concrete syntax (actually, it is a concrete syntax meta-model). We need parser to construct AST so we use xText grammar syntax as this framework solves the problem in most elegant way. We can automatically generate initial grammar sample (representing primitive syntax) to be edited by user then.

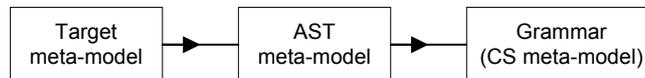

**Fig. 1.** Defining meta-models

The last stage defines text-to-AST transformation (as xText does). AST-to-model transformation skeleton is implicitly defined on the second stage. Thus, this completes text-to-model transformation.

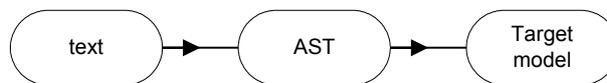

**Fig. 2.** Text-to-model transformation

Since text-to-AST transformations are successfully handled by xText, we concentrate on AST-to-model transformations.

# 3 Transformation actions

As mentioned above AST-to-model transformation skeleton is implicitly defined when transforming target meta-model to AST meta-model. Here we examine what actions are usually held when transformation target meta-mode to AST meta-model. These actions are analyzed to define a transformation language.

## 3.1 Structure

AST classes usually have the same overall inheritance structure and aggregation associations so we can initially map each target class to an AST class with the same structural features (these features are going to be slightly modified, see below).

When we have the following Class class in target meta-model:

**class** Class **extends** Classifier {

    **ref** Class[*] super;

    **attr boolean** abstract;

    **val** StructuralFeature[*] structuralFeatures;

}

The corresponding AST class will appear like this:

**class** ClassAS **extends** ClassifierAS {

    **ref** ClassAS[*] super;

    **attr boolean** abstract;

    **val** StructuralFeatureAS[*] structuralFeatures;

}

The only problem here is that the cross-reference *super* does not actually exist in the AST, but a *String attribute* exists instead (see the next subsection).

## 3.2 References

The most interesting thing is to translate references. In the target meta-model, a Class references its superclass by a structural feature of type Class and in the AST the same reference is done by a String.

Thus, we need to define the translation saying that within the AST Class is (cross-)referenced using String. Such translated references will be resolved to model ones through lookup actions during AST-to-model transformation.

**Reference kinds.** In our example above, we change the type of cross-reference *super* but containment-reference *structuralFeatures* remains unchanged. This is right in assumption of all the containment in target model being somehow containment in AST (and represented by rule references in grammar). This is natural for most languages, but might be a drawback in the general case.

Thus, we might need to define that some references within the AST are no longer containment (actually we mention it here only for completeness, our transformation language do not allow this). In addition, we never change types of containment references.

### 3.3 Syntax-specific classes

Sometimes AST meta-model must have some classes which have no prototypes in target meta-model.

**Compilation unit.** The most obvious example is a compilation unit class. This class represents a root of AST which is connected to file-based distribution of compilation units (files containing code) and usually has no corresponding class in target meta-model. Thus, we must create this class within the AST meta-model and transform it to nothing during AST-to-model transformation.

**Qualified names.** Another example is a qualified name: this class is extremely syntax-specific: as it denotes just a structured string reference, it cannot have any prototype in target meta-model. Thus, we create it too and transform to a reference in target model (because some class will be referenced by qualified name).

Inference: we might need to create an AST class without a prototype. Object of such classes are not transformed to target model directly.

### 3.4 Inheritance

AST classes' inheritance is based on syntactical generalization and target classes' inheritance is based on domain notions' generalization. Thus, inheritance hierarchy might be changed in AST meta-model in comparison to target meta-model.

This might occur when some classes in target meta-model are enumerations or singletons. Almost no modeling tool allows treating such classifiers as classes. They are treated as primitive types. Possibly, it is right because objects of these classes cannot be contained by other objects.

Prevailing example is modeling basic type (like in Java): there are only eight basic types, so they constitute an enumeration, but generalization hierarchy of *type* notion requires polymorphism between classes and basic types. Thus, new reference-class is represented within the meta-model:

```
class BasicType extends Type {
    attr BasicTypeEnum type;
}
```

Objects of this class represent references (which are usually containment-references) to basic types. When we transform such a target meta-model to an AST meta-model and translate references to classes to be qualified names, we get the problem: a qualified name represents a type and the BasicTypeAS (class created from BasicType as a prototype) represents a type too, thus, they must have a common superclass. QualifiedName must be descendant of TypeAS, but packages are referenced by qualified names too, and one cannot specify a basic type reference where the package must appear. Solution is the following:

1. Create a new AST class ClassReferenceAS containing a qualified name :
   ```
   class ClassReferenceAS extends TypeAS {
       val QualifiedName name;
   }
   ```
2. Make ClassAS not inherit from TypeAS any more since TypeAS now represents type *reference* not definition.

Inference: we must be able to change inheritance in the AST meta-model. This does not affect AST-to-model transformation anyhow.

### 3.5 Non-syntax classes

Some target classes might have no representation in concrete syntax and thus, in AST.

The best example is a package. In a target meta-model Package aggregates classes and subpackages but in concrete syntax this notion is represented by directory structure. Thus, we do not have any PackageAS in AST meta-model, but we can have some kind of reference to a Package (qualified name) being resolved during AST-to-model transformation.

We need to skip some classes in target meta-model and do not transform them into AST classes. This does not affect AST-to-model transformation except for the fact that objects of such classes must be created manually (no code can be generated for creating them).

## 4 Transformation language

In the previous section, we have outlined some actions taken when transforming target meta-model to AST meta-model. Here we describe corresponding language – a DSL for such transformations. This is done according to the proposed approach: starting with target meta-model and then proceeding to concrete syntax. Additionally, our language is used to describe itself.

### 4.1 Target meta-model

Here we define classes for actions described above and some formalized semantics for these actions.

**Common superclass.** All these classes are descendants of the common abstract superclass Action.

```
abstract class Action {
}
```

**Class mapping.** The implicitly mentioned element is a class mapping. This element connects a prototype class in target meta-model with its AST image.

```
class ClassMapping extends Action {
    ref EClass prototype;
    ref EClass image;
}
```

*Semantics.* Instances of this class will be initially created for each target meta-model class. For such instance, the *prototype* is that target class and the image is build according to the following rules:
- it is abstract if and only if the *prototype* is abstract;
- its name is the *prototype* name suffixed with 'AS';
- its list of superclasses consists of images of the *prototype's* superclasses;
- it has the same number of structural features with the same names as the *prototype* has;
- each attribute has the same type when the corresponding *prototype's* attribute has;
- each reference is of type that is image for the class type the corresponding *prototype's* reference has.

**Translate references.** This is an action denoting that some reference has some specific textual representation.

```
class TranslateReferences extends Action {
    ref EClass modelReferenceType;
    ref EClassifier textualReferenceType;
    attr boolean includeDescendants;
}
```

*Semantics.* All the cross-references of type *modelReferenceType* (and all of its subtypes if *includeDescendants* is true) in AST classes change type to *textualReferenceType* and become containment. If *textualReferenceType* is EDataType those references are replaced with attributes.

Usually, all the stored (not *derived*, in EMF terms) cross-references must be translated to be represented textually.

**Class creation.** This action creates a (syntax-specific) class.

```
class CreateClass extends Action {
    attr String name;
    attr boolean abstract;
    ref EClass[*] superclasses;
    val StructuralFeature[*] structuralFeatures;
```

```
    }
    abstract class StructuralFeature {
        attr String name;
        attr int lowerBound;
        attr int upperBound = 1;
    }
    class Attribute extends StructuralFeature {
        ref EDataType type;
    }
    class Reference extends StructuralFeature {
        ref EClass type;
        attr boolean containment;
    }
```

*Semantics.* An AST class with specified properties is created (*StructuredFeature* and its descendants are mapped to corresponding ECore classes).

**Changing inheritance.** These actions change inheritance structure.

```
    class ChangeInheritance extends Action {
        ref EClass target;
        ref EClass[*] superclasses;
    }
```

*Semantics. Target's* list of superclasses is set to *superclasses* value.

**Handling non-syntax classes.** This action prevents some target meta-model class to have an image.

```
    class SkipClass extends Action {
        ref EClass target;
        attr boolean includeDescendants;
    }
```

*Semantics. Target's* image is removed from AST meta-model. All of its subtypes' images are removed too if *includeDescendants* is true. If there are some cross-references to removed classes and there are actions to translate these references then these actions must perform correctly regardless to being executed before or after *SkipClass* action.

**Transformation element.** This utilitarian class serves as a container for action sequence.

```
    class Transformation {
```

```
    val Action[*] actions;
}
```

*Semantics.* All the *actions* are executed (execution order is not defined).

### 4.2 AST

Since AST is defined by transforming target meta-model with the language we are defining the AST for, we provide the transformation in our concrete syntax with necessary annotations.

At first, create a new class for qualified names. The syntax is Emfatic-like except for **create** keyword.

```
create class QualifiedName {
    attr String name;
    val QualifiedName subQN;
}
```

Then translate references to image of EClassifier and its subtypes to be represented with QualifiedNames.

```
refer img(ecore::EClassifier)+ as QualifiedName;
```

The last thing: EClassifier and its subtypes and ClassMapping have no image.

```
skip ecore::EClassifier+;
```

```
skip ClassMapping;
```

That is all since our language is rather simple.
We get the following AST (a GMF diagram is provided for short):

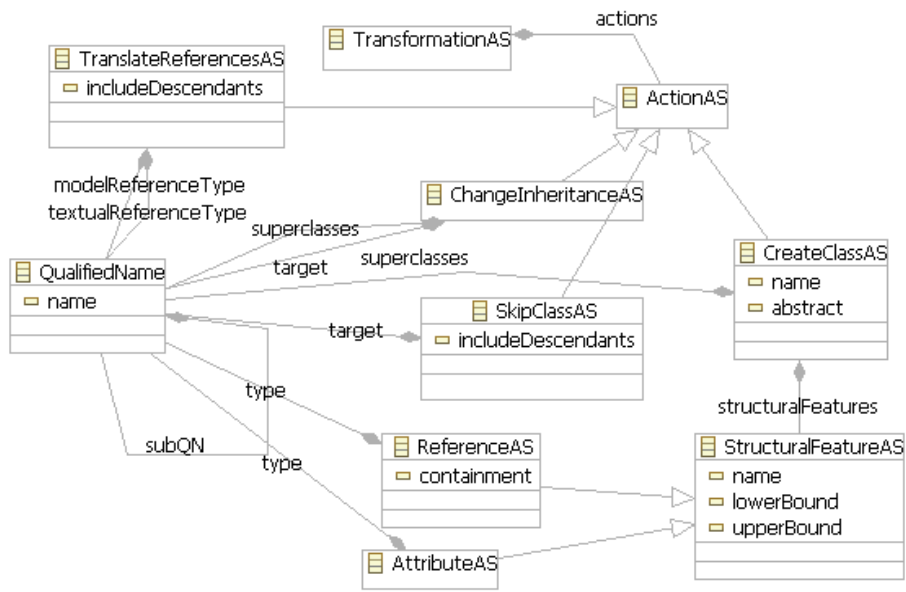

**Fig. 3.** Transformation language AST meta-model

### 4.3 Concrete syntax (rule samples)

Here are sample rules from xText grammar for our language. The whole grammar appears to be too large and not so interesting.

At first, the following rule describes syntax for ChangeInheritanceAS which was not used above.

```
ChangeInheritanceAS:
  "make" "img" "(" target=QualifiedName ")" "extend"
    ("nothing" | (superclasses+=QualifiedName
      ("," superclasses+=QualifiedName)*)?);
```

Another example: an abstract rule for StructuralFeatureAS, subclasses are represented as alternatives.

```
Abstract StructuralFeatureAS:
  AttributeAS | ReferenceAS;
```

In addition, this describes syntax for CreateClassAS: **abstract** keyword is an optional modifier here.

```
CreateClassAS:
  "create" (abstract?"abstract") "class" name=ID
    ("extends" superclasses+=QualifiedName
      ("," superclasses+=QualifiedName)*)? "{"
    (structuralFeatures+=StructuralFeatureAS ";")*
  "}";
```

### 4.4 Language summary

A simple transformation language that we have defined gives rather good proof of our approach applicability. This language is strong enough to define itself (and many other domain-specific languages like Emfatic or CSS).

Our approach is to construct the language starting with its target meta-model (considering its semantics). This approach makes developer concentrate on domain notions, not syntax. Such a workflow if much more language engineering style that a syntax-based one is.

## 5 AST-to-model transformation

We have described out meta-model transformation language in the previous section. Despite the fact it operates on meta-models, this language can be used to generate a skeleton of AST-to-model transformation reversing transformation direction and holding it one meta-level lower.

### 5.1 Transformation structure

We use meta-model transformation data to transform AST to target meta-model.

The most obvious thing is to look at ClassMapping actions and for each *image* instance to create a *prototype* instance copying all the attributes and transforming containment.

Thus, TransformationAS instance will be transformed to Transformation instance and all the ActionAS instances it contains will be transformed to corresponding prototype instances. For example, each SkipClassAS instance will be transformed to SkipClass and *includeDescendants* value will be simply copied.

### 5.2 Name lookup

The only problem is to retrieve right values for target model's cross-references. Since we changed reference types for cross-references and they are represented textually (by names, qualified names, numbers or whatever), we must have some actions to reconstruct model references from those textual representations.

These actions might be instituted as methods of a helper class written manually (in Java, since we use Java-based tools).

These methods will have the following form:

```
TargetModelClass lookupReference(ASTClass textualRef) {
    // ...
}
```

They might be called by generated code to resolve cross-reference values. The helper class containing all such methods will define lookup semantics for the whole language.

It might be better to split this single class into separate classes for each reference type since it increases modularity, but in practice, this appears to be too much classes having one single line method inside.

### 5.4 Trace data

To generate such lookup methods we need some information about which reference in AST meta-model corresponds to which one in target meta-model. This is handled by meta-model transformation *trace* (see [10] for strict definition) that records all the actions with sources and targets providing all the information necessary to reverse the transformation.

Trace records are described by a separate meta-model that references our language's target meta-model.

## 6 Conclusion

Model-driven and generative technologies get stronger with time passing and domain-specific languages become more reachable and, hence, useful [11, 12]. Although there are different non-textual syntax representations, we believe that textual languages are still needed and, sometime, irreplaceable.

In this paper, we made another step towards simplified textual DSL development process through formalizing the semantic analysis phase and making the target meta-model the first and main artifact. This is done by using rather simple language and generating transformations (and by using *xText*, of course).

The system described by this page is currently under development on the very early phase. We plan to finish implementing the key features by the April 2007.

In the following subsections, we denote some directions of further research and development we plan ourselves and welcome others to participate.

### 6.1 Lookup data structures and modeling

Currently no data structures used during lookup process (symbol tables or other structures) are provided by the framework itself but might be chosen on a free basis. This allows developers to use the most suitable (simple and efficient) data structures they need.

Anyway, some standard framework might be developed to handle most frequently appearing tasks (like hierarchical namespaces and stubs for forward declarations) or provide ready-to-use solutions, for example, to lookup EClass'es by qualified names. Probably, we can even model some (simple) lookup cases but this topic needs some more research activities.

## 6.2 AST-to-text transformation

When you use some models in a collaborative development, some issues appear with comparing and merging changes with a repository. Communities develop tools to be able to compare models in their graphical concrete syntaxes, but actually, this is very complicated especially when we have some homegrown syntax and want to merge, not only compare. Of course, we can try to generate a compare/merge tool along with the editors using the same framework, but this needs a lot of additional information and makes users understand complicated comparison semantics.

On the other hand, everyone is used to textual compare/merge tools presented in every version control system. It is much easier to store your models in textual form (not XML/XMI with their unreadable references but comprehensible declarative DSL form like Emfatic) and compare/merge them textually than to write (or generate) your own model compare tool.

Therefore, we need a simple way to store models textually, even if we edit them in some graphical syntax. This means we need to generate a textual representation of a model automatically. In our framework this just means to transform a model to the corresponding AST (almost the same way the AST is transformed to the model) and then transform this AST to text (here we need to build a text-generator from xText grammar).

This seems to be rather useful and easy to implement and is the first topic on our further research list.

## 6.3 Error handling and validation

Some errors might occur during translation. For example, some names might stay unresolved or some target meta-model constraints might not be met. Additionally, some syntax or lexical errors may occur. All these problems must be detected and reported.

Currently we use EMF Validation Framework [13] to handle this but it handles only model constraints and does not integrate with other error sources.

We plan to develop some common problem-reporting framework and maybe define some model extensions to generate some checkers.